\documentclass[aps,twocolumn,showpacs,preprintnumbers]{revtex4}

\usepackage{graphicx}  
\usepackage{subfigure}
\usepackage{multirow}
\usepackage{verbatimbox}
\usepackage{fancyhdr}
\usepackage{longtable}
\usepackage{parskip}
\usepackage[T1]{fontenc}
\usepackage{dcolumn}   
\usepackage[title]{appendix}
\usepackage{float}

\usepackage{bm}        
\usepackage{amsfonts}  
\usepackage{amsmath}   
\usepackage{amssymb}   
\usepackage{titlesec} 
\usepackage{xcolor}

\newcommand{\pwisein}{\left\{ \begin{array}{ll}}
\newcommand{\pwiseout}{\end{array}\right.}

\def\nicefrac#1#2{%
    \raise.5ex\hbox{#1}%
    \kern-.1em/\kern-.15em%
    \lower.25ex\hbox{#2}}

\titlespacing{\subsection}{0pt}{5pt}{5pt}

\begin{document}

\title{Numerical scheme for the far-out-of-equilibrium time-dependent Boltzmann collision operator: 1D second-degree momentum discretisation and adaptive time stepping}

\author{I.~Wadgaonkar}
\affiliation {Nanyang Technological University, 21 Nanyang Link, Singapore, Singapore}
\author{R.~Jain}
\affiliation {Nanyang Technological University, 21 Nanyang Link, Singapore, Singapore}
\author{M.~Battiato}
\email{marco.battiato@ntu.edu.sg}
\affiliation {Nanyang Technological University, 21 Nanyang Link, Singapore, Singapore}

\date{July 1 2020}

\begin{abstract}  

Study of far-from-equilibrium thermalization dynamics in quantum materials, including the dynamics of different types of quasiparticles, is becoming increasingly crucial. However, the inherent complexity of either the full quantum mechanical treatment or the solution of the scattering integral in the Boltzmann approach, has significantly limited the progress in this domain.
In our previous work we had developed a  solver to calculate the scattering integral in the Boltzmann equation. The solver is free of any approximation (no linearisation of the scattering operator, no close-to-equilibrium approximation, full non-analytic dispersions, full account of Pauli factors, and no limit to low order scattering) \cite{Michael}. Here we extend it to achieve a higher order momentum space convergence by extending to second degree basis functions.
We further use an adaptive time stepper, achieving a significant improvement in the numerical performance. Moreover we show adaptive time stepping can prevent intrinsic instabilities in the time propagation of the Boltzmann scattering operator.
This work makes the numerical time propagation of the full Boltzmann scattering operator efficient, stable and minimally reliant on human supervision.

\end{abstract}

\maketitle 

\section{Introduction}

The study of ultrafast dynamics\cite{Fujimoto1984,Elsayed1987,Schoenlein1987,Brorson1990,Fann1992,Hertel1996,Fatti2000,Stamm2007} has allowed for the discovery of a number of very interesting effects\cite{Beaurepaire1996,Terada2008,Crepaldi2012,Garcia2011,Cacho2015,Battiato2018PRLHalfMetals,Cheng2019,Wang2018THzTopolInsul,Lindemann2019UltrafastSpinLasers}. Its theoretical description however has to face a number of challenges, since the most interesting effects often come from the interplay of laser excitation, thermalisation of different types of quasiparticles and transport\cite{malinowski2008control, battiato2010superdiffusive,rudolf2012ultrafast,kampfrath2013terahertz, eschenlohr2013ultrafast, battiato2016ultrafast,Freyse2018}. The concomitant description of these different effects is further aggravated by the fact that out-of-equilibrium dynamics dramatically increases the complexity, as the population cannot be anymore safely assumed as close to equilibrium\cite{Bagsican2020THzCNT}. The large amount of degrees of freedom required to describe a band, position and momentum dependent population increase vastly any theoretical and numerical effort. It is also evident that at times the need to describe several types of quasiparticles requires input from a number of methods, further increasing the complexity. Finally, all these data about transport and scattering for different quasiparticles need to be integrated into a single treatment.

The time dependent Boltzmann equation (BE)\cite{Boltzmann1872} has proven very powerful in tackling this complexity as it allows for a seamless integration of transport and scattering even among different types of quasiparticles \cite{Semiconductor1969,cercignani1988boltzmann,Abdallah1996,Mahan2000,Choquet2000,Rethfield2002,Majorana2004,Caceres2006,Snoke2011,Tani2012,Vatsal2017,wais2018quantum}. This approach requires the tackling of two fundamental issues. First, the BE needs \textit{ab initio} input for the dispersions and the scattering matrix elements. While finding dispersions for quasiparticles is nowadays routine work in many cases, the second part is far more challenging, yet recently it has been successfully tackled for different quasiparticles by several groups \cite{jhalani2017ultrafast,li2014shengbte,lindsay2014phonon} (we do not address this issue here). The second issue, which we address in this work, is the discretisation of the time-dependent BE itself, after the \textit{ab initio} input is known.

The BE has been widely used in a range of fields spanning from gasses, plasma, and semiconductor's physics (we focus here specifically on the time-dependent equation)\cite{colonna2016plasma,villani2002review,kremer2010introduction,saint2009hydrodynamic,colussi2015undamped,snoke2011quantum,shomali2017monte,xu2017lattice}. The transport part has a similar shape in all those fields, with the only notable exception that in solid state applications, the particle's dispersion is not anymore a quadratic function of the momentum, but is in general much more complicated and often an analytic form is not known. A number of successful numerical methods to tackle the transport part of the BE have been developed in the past, even in the presence of electric or electromagnetic fields\cite{morgan1990elendif,nabovati2011lattice,sellan2010cross,hamian2015finite,romano2015dsmc,heath2012discontinuous,cockburn2001runge,li1998analytic,choquet2003energy,majorana2004charge,Singh2020Boltztransp,Singh2020Boltztransp2}.

The scattering part has instead always been more challenging \cite{vangessel2018review,chernatynskiy2010evaluation,broido2007intrinsic,wu2013deterministic,bird1994molecular,homolle2007low,tcheremissine2006solution,ibragimov2002numerical,gamba2009spectral,pareschi2000numerical,mouhot2006fast,tani2012ultrafast,maldonado2017theory,ziman2001electrons,fischetti1988monte}.
 For classical particles (gasses and plasmas) as well as in the low occupation regime in semiconductors, the scattering term still requires handling high dimensional integrals, but it can be written as a linear operator of the particles' populations. A number of approximations have been used to address this problem, yet with modern computational capabilities the linear problem can be handled. On the other hand, when the quantum statistics of particles start playing a role, the scattering term of the BE becomes, in the case, for instance, of electron-electron collisions, a quartic operator due to the presence of the Pauli factors. This hugely increases the difficulty of addressing this problem with straightforward numerical approaches. It is very common in these cases to do close to equilibrium approximations to again reduce the collision term to a linear or at most quadratic operator.

All the mentioned difficulties are present in steady state calculations, but computing the time evolution aggravates some of them and rises further numerical challenges. The scattering integral (which we remind is a quartic operator) requires high dimensional integrals, with highly discontinuous integrands. Yet, even more importantly, the issue of particle, energy and momentum conservation becomes critical. A numerical method that does not preserve exactly those quantities can be used reliably only for steady state or short time propagations, making exact conservation an indispensable property. Unfortunately standard ways of calculating the scattering integrals lead to errors that break such conservation laws.

We have proposed a numerical method to solve the BE scattering term, which allowed for the treatment of arbitrary dispersions, arbitrary scattering amplitudes, had an excellent scaling, and preserved particle, energy and momentum\cite{Michael,Bagsican2020THzCNT}. We here extend that method to a higher order and integrate an adaptive time stepper, vastly increasing the numerical performance.

\section{The Boltzmann scattering integral}

The scattering part of the BE (a.k.a.~the quantum Fokker-Planck equation) can be obtained by time-dependent perturbation theory applied to a hamiltonian where any number of quasiparticles weakly interact with each other\cite{snoke2011quantum,snoke2020solid}. Notice that the BE cannot be applied to strongly interacting particles directly, as time-dependent perturbation theory fails. In that case one should first reformulate the problem by finding weakly interacting quasiparticles.

Assuming that all quasiparticles' dispersion $\epsilon_n\left( \bold{k}\right)$ (where the index $n$ is a composite index containing the quasiparticle type and the band index, and $\bold{k}$ the crystal momentum) are known, the scattering BE provides the expression for the time ($t$) evolution of the momentum resolved population $f_n\left(t,\bold{k}\right)$ for each of the band for each quasiparticle involved in the scattering.
Each combination of scattering among quasiparticles and bands with an active interaction (for shortness called scattering channel) will contribute to the time evolution. 

\begin{figure}
    \centering
    \includegraphics[scale=0.3]{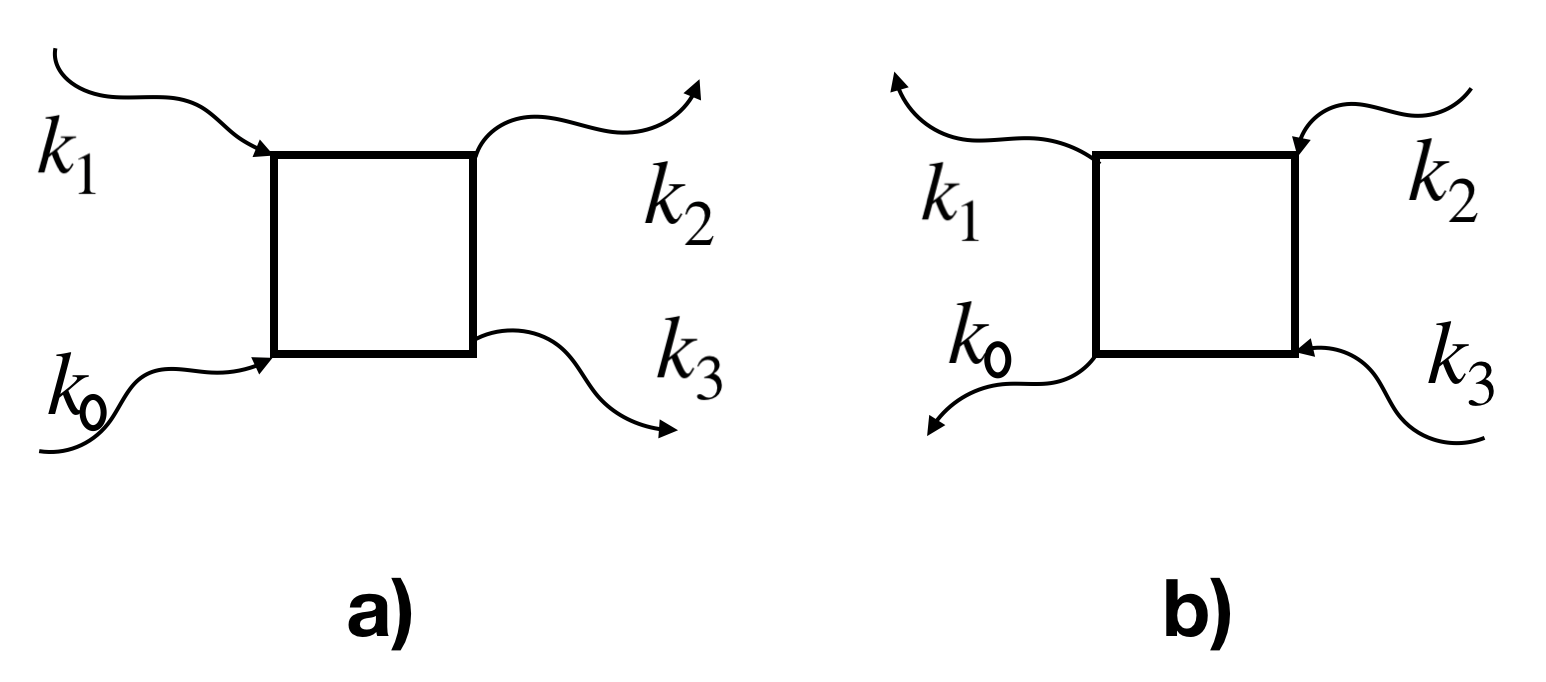}
    \caption{a) Electron-electron scattering between an electron in a state labelled by $k_0$, $n_0$ and another in a state labelled by $k_1$, $n_1$ to states labelled by $k_2$, $n_2$ and $k_3$, $n_3$. b) Time-reversed process corresponding to the same scattering.}
    \label{fig:Scattering Fenyman's diagram}
\end{figure}

As an example, the Boltzmann scattering term (BST) for a four fermionic legs scattering $n_0 + n_1 \leftrightarrow n_2 + n_3$ (which, for simplicity, we suppose different) is composed of four terms giving the time propagation of the population in each involved leg. The first term, giving the time evolution of the population in the first involved band $n_0$, is
\begin{widetext}
\begin{equation}\label{ScatteringIntegral}
\begin{split}
\Bigg(\frac{\partial f_{n_0}\left(t,\bold{k}_0\right)}{\partial t}\Bigg)_{ \substack{ n_0 + n_1 \\  \leftrightarrow \\ n_2 + n_3} }
=  \sum_{\bold{G}} \int \int \int_{V^3_{BZ}}  & d\bold{k}_1 \, d\bold{k}_2 \, d\bold{k}_3 \;\; w_{ \substack{ n_0 + n_1 \\  \leftrightarrow \\ n_2 + n_3} } \left( \bold{k}_0,\bold{k}_1,\bold{k}_2, \bold{k}_3\right)\; \cdot \\ 
& \delta(\epsilon_{n_0}(\bold{k}_0)+\epsilon_{n_1}(\bold{k}_1)-\epsilon_{n_2}(\bold{k}_2)-\epsilon_{n_3}(\bold{k}_3)) \;\; \delta(\bold{k}_0+\bold{k}_1-\bold{k}_2-\bold{k}_3+\bold{G}) \cdot \\ 
&  \Big[\big(1-f_{n_0}\left(t,\bold{k}_0\right)\big)\big(1-f_{n_1}\left(t,\bold{k}_1\right)\big) f_{n_2}\left(t,\bold{k}_2\right) f_{n_3}\left(t,\bold{k}_3\right)  \\
& \;\;\; -f_{n_0}\left(t,\bold{k}_0\right) f_{n_1}\left(t,\bold{k}_1\right) \big(1-f_{n_2}\left(t,\bold{k}_2\right)\big) \big(1-f_{n_3}\left(t,\bold{k}_3\right)\big) \Big], 
\end{split}
\end{equation}
\end{widetext}
where $\sum_\bold{G}$ denotes summation over all reciprocal lattice vectors $\bold{G}$ to account for umklapp scattering and $w$ is the momenta ($\bold{k}_n$) dependent scattering amplitude (which is supposed to be known or estimated by other means) for that given scattering channel. The triple integral on the momenta over the volume of Brillouin zone $V_{BZ}$, along with the two Dirac deltas accounts for all combinations of momenta  which yet have to satisfy energy and momentum conservation (up to a reciprocal lattice vector $\bold{G}$). The two addends in the square brackets, which we call scattering phase space, account for the direct and the time reversed processes (see Fig.~\ref{fig:Scattering Fenyman's diagram}). 

The generalisation of Eq.~\eqref{ScatteringIntegral} to other types of scatterings between different quasi-particles (both fermions and bosons) or even different number of legs (a.k.a.~involved states) keeps the mathematical structure unchanged. For this reason we will show the numerical method for the time propagation in Eq.~\eqref{ScatteringIntegral} only, yet it is general to any type of scattering. Let us, however, remind that calculating quasiparticle dispersions $\epsilon_n\left( \bold{k}\right)$ and the momenta-resolved amplitudes $w\left( \bold{k}_0,\bold{k}_1,\bold{k}_2, \bold{k}_3\right)$, which are required as input in Eq.~\eqref{ScatteringIntegral}, are not addressed here. 

Assuming all the input functions are known, the numerical treatment of Eq.~\eqref{ScatteringIntegral} still presents several critical challenges. 1) The integral in Eq.~\eqref{ScatteringIntegral} is high dimensional and it contains several (depending on the dimensionality of the system) Dirac deltas, one of which has a highly non trivial form. This means that the scattering term requires the execution of integrals over an extremely complex high-dimensional hyper-surface. 2) As all the populations appearing in Eq.~\eqref{ScatteringIntegral} are time dependent and known only at runtime, the integral operator on the right-hand side is a quartic operator. This leads to a very prohibitive scaling of the storage and computational cost, if straightforward methods are used. 3) It is imperative to construct a numerical method that is able to conserve exactly particle number, total momentum and total energy, regardless of the achieved numerical precision. In the absence of such property, any numerical method can be exploited only for limited time propagations or under some strict conditions and approximations.

In the following sections we will address how to extend the method we developed\cite{Michael} in two major ways: 1) the increase of the order of convergence, by using second degree piecewise continuous  polynomial basis functions and 2) the use of a more advanced, higher order time stepping algorithm with adaptive timestep.

\section{Momentum space discretization}

We present in this section the necessary changes to our method\cite{Michael} in order to increase the momentum space order of convergence. We restrict ourselves in this work to one dimensional materials, yet it can be generalised to higher dimensions. For the reader's convenience we will briefly summarise the structure of the method, which is independent of the basis functions. We will then introduce the second degree basis functions used in this work. Then we will show how the construction of the scattering tensor has to be modified to account for the use of second degree polynomial basis functions, and how to construct the scattering elements. 

We assume the distribution functions $f_{n}\left(t,k\right)$ and the dispersion relation $\epsilon_{n}\left(k\right)$ for a given band $n$ are defined over a certain domain, which can be a compact subset of the Brillouin zone. The method will work even when a different domain is chosen for each band, allowing computational savings as certain regions of the Brillouin zone might not be affected by the dynamics and can be efficiently disregarded. 

We now focus on a single band. We split the domain into non-overlapping regions called 'elements' which completely cover that domain. Together, the group of elements is called a mesh. For our 1D analysis the mesh consists of divisions of the line. Again every band can have a different mesh, and even if in the present work the results are presented for uniform meshes, the method can be applied to any unstructured mesh.

\subsection{Projection on local basis functions}

We project all the functions defined over the domain of a given band $n$ on a set of momentum basis functions $\Psi_{\substack{A a\\ n}} (k)$ (which again can be different for each band). We construct these basis functions to be zero everywhere except over the element identified by the index $A$, and continuous everywhere except at the edges of the element $A$. More than one basis function with subindex $a$ can be constructed for each element as long as they are linearly independent. For instance, the distribution functions  $f_{n}\left(t,k\right)$ and the dispersion relation $\epsilon_{n}\left(k\right)$ can be written as a linear combination of these basis functions:
\begin{equation}\label{fandEProjection}
\begin{split}
f_n(t,k)&=\sum_{B b} f_{\substack{B b \\ n}}(t) \; \Psi_{\substack{B b \\ n}} (k), \\
\epsilon_n(k)&=\sum_{B b} \epsilon_{\substack{B b \\ n}} \; \Psi_{\substack{B b \\ n}} (k),
\end{split}
\end{equation}
where $f_{\substack{B b \\ n}}(t)$ and $\epsilon_{\substack{B b \\ n}}$ are the coefficients of the discretised population and dispersion. Notice that $f_n(t,k)$ has been so far only semi-discretized since the time variable in the coefficients $f_{\substack{B b \\ n}}(t)$  has not been discretized yet. 

We further assume here (yet it is not necessary) that our basis functions are orthonormal:
\begin{equation}
	\int dk \; \Psi_{\substack{A a \\ n}} (k)\; \Psi_{\substack{B b \\ n}} (k)  = \delta_{AB} \;\delta_{ab} 
\end{equation}
where $\delta$ is the Kronecker delta.  Projecting the collision integral in Eq.~\eqref{ScatteringIntegral} on the chosen basis and using Eq.~\eqref{fandEProjection}, the orthonormality of basis functions, and the fact that the basis functions are non zero only over a single element, we obtain the final expression for the semi-discretised form of collision integral as (see ref.\cite{Michael} for a detailed derivation)
\begin{widetext}
\begin{equation}\label{CollisionIntegral}
\begin{split}
\sum_{A a'} \frac{d f_{\substack{A a' \\n_0}}(t)}{d t}=\sum_{A a} \sum_{B b} \sum_{Cc} \sum_{Dd} S^{a' a b c d }_{\substack{ABCD \\ n_0 n_1 n_2 n_3}}  \Bigg(&\left(1_{\substack{Aa \\ n_0}}- f_{\substack{Aa \\n_0}}(t)\right) \left(1_{\substack{Bb \\ n_1}}- f_{\substack{Bb \\n_1}}(t)\right)  f_{\substack{Cc \\n_2}}(t)\;f_{\substack{Dd \\n_3}}(t) -\\
& f_{\substack{Aa \\n_0}}(t)\; f_{\substack{Bb \\n_1}}(t) \left(1_{\substack{Cc \\ n_2}}-f_{\substack{Cc\\n_2}}(t)\right) \left(1_{\substack{Dd \\ n_3}}-f_{\substack{Dd \\n_3}}(t)\right) \Bigg) \\
\end{split}
\end{equation}
with
\begin{equation}\label{ScatteringTensor}
\begin{split}
S^{a' a b c d }_{\substack{A B C D \\ n_0 n_1 n_2 n_3}}=\sum_G \int\displaylimits_{\substack{A \\ n_0}} \int\displaylimits_{\substack{B\\ n_1}} \int\displaylimits_{\substack{C \\ n_2}} \int\displaylimits_{\substack{D \\ n_3}} &  dk_0 \,dk_1\, dk_2\, dk_3  \;\; 
\Psi_{\substack{Aa' \\ n_0}} (k_0)\Psi_{\substack{Aa \\ n_0}} (k_0)\Psi_{\substack{Bb \\ n_1}} (k_1)\Psi_{\substack{Cc \\ n_2}} (k_2)\Psi_{\substack{Dd \\ n_3}} (k_3)  \\ 
& w_{ \substack{ n_0 + n_1 \\  \leftrightarrow \\ n_2 + n_3} } \left( k_0,k_1,k_2, k_3\right) \;\; \delta(k_0+k_1-k_2-k_3+G)\\
& \delta\Big(\sum_{\alpha} \epsilon_{\substack{A\alpha \\ n_0}} \Psi_{\substack{A\alpha \\ n_0}} (k_0)+\sum_{\beta} \epsilon_{\substack{B \beta \\ n_1}} \Psi_{\substack{B \beta \\ n_1}} (k_1)-\sum_{\gamma} \epsilon_{\substack{C \gamma \\ n_2}} \Psi_{\substack{C \gamma \\ n_2}} (k_2)-\sum_{\delta} \epsilon_{\substack{D \delta \\ n_3}} \Psi_{\substack{D \delta \\ n_3}} (k_3) \Big)
\end{split}
\end{equation}
\end{widetext}
where $1_{\substack{Aa \\ n}}$ is the discretized representation of a function with a constant value of 1 over the domain, and the integrals are now only over a single element and not anymore over the full domain. We refer to $S^{a' a b c d}_{\substack{A B C D \\ n_0 n_1 n_2 n_3}}$ as the scattering tensor, since it contains all information about the scattering. Even if the numerical results shown in this work are obtained using orthonormal basis functions, we stress that the orthonormality assumption is not necessary and the method is immediately generalisable to a non-orthonormal basis set by simply adding a mass matrix to Eq.~\ref{CollisionIntegral}. We will not repeat here how to execute the summation in Eq.~\ref{CollisionIntegral} as it is already described in Ref.~\cite{Michael}, and we will simply focus on the construction of the scattering integrals in Eq.~\ref{ScatteringTensor}.

\subsection{Second degree polynomial basis functions}

\begin{figure} [tb]
    \centering
    \includegraphics[scale=0.4]{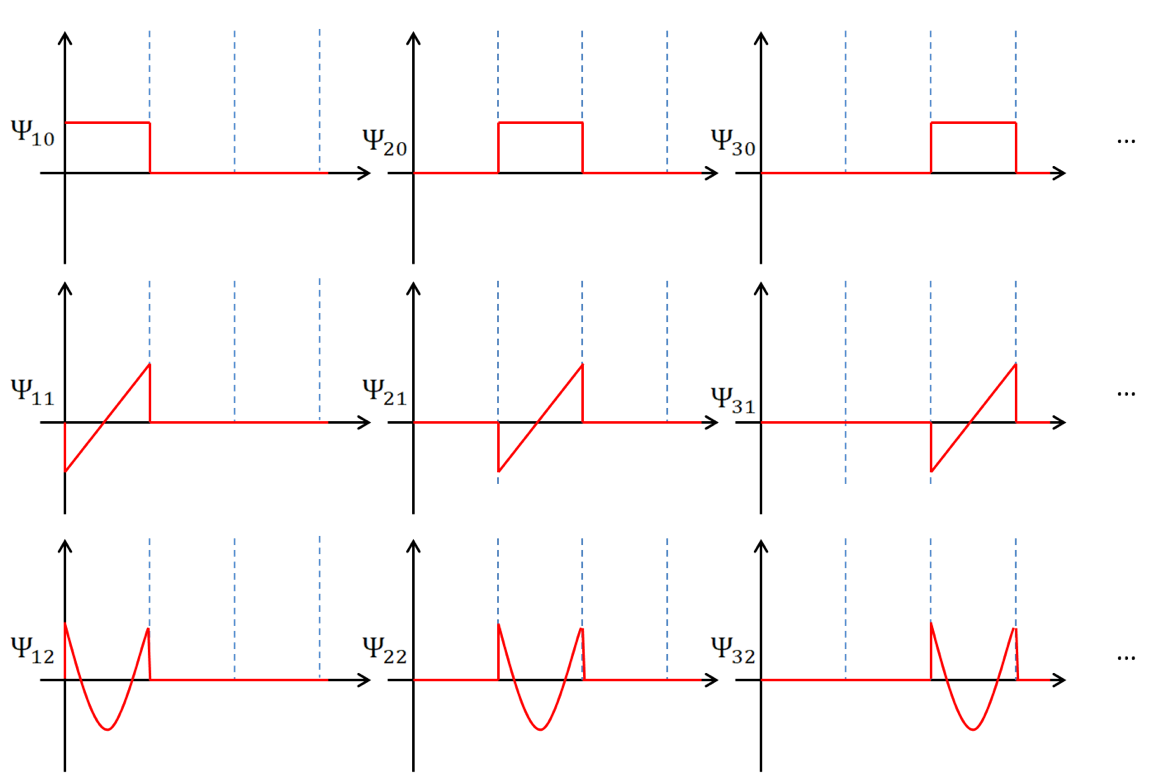}
    \caption {Example of basis functions for a given band. The band index is omitted for brevity.}
    \label{fig:BasisFunctions}
\end{figure}

To achieve a higher order of convergence compared to our previous work, we now use as basis functions second degree piecewise discontinuous polynomials with support on only one element, as shown in Fig.~\ref{fig:BasisFunctions}. We choose normalized Legendre polynomials of order 0,1 and 2 as basis functions for each element since they form an orthonormal set. The scaling of the number of non-zero elements of each scattering channel remains the same as in the case of linear basis functions. 
However, switching from first degree to second degree basis functions has three main effects. a) The time and storage cost increases in 1D and for 4-leg scatterings by a factor $(3/2)^5\approx 7.6$, as there are more basis functions and therefore more $(a'abcd)$ combinations for each combination of $(ABCD)$. b) The order of convergence is however one order higher than the linear case (as it scales as $\epsilon \sim (1/N )^{(p+1)}$, where p is the maximum order of basis functions and N is the number of elements in the mesh), and, already at very low precisions, it compensates for the cost increase and outperforms the lower order approach. c) The third difference is however the most critical: the energy Dirac delta has now a quadratic function inside. This leads to a large increase in cost compared to the linear order in the pre-calculation, yet interestingly it does not affect the storage cost and the numerical cost of the time propagation.

\subsection{Numerical integration of scattering tensor elements}

The presence of Dirac deltas for energy and momentum conservation in Eq.~\eqref{CollisionIntegral}, makes the integration domain highly discontinuous thereby making the use of a straightforward Monte Carlo integration approach impractical. However, notice that our choice of basis functions transforms the dispersion into a piecewise polynomial as shown in Eq.~\eqref{ScatteringTensor}. As a result we can analytically invert both the Dirac deltas to reduce the integrand to a more manageable form. We now need to choose the variables to invert the Dirac deltas. Although all the choices are analytically equivalent, they are not so numerically. Moreover, each inversion will lead to slightly different formulas, due to sign asymmetry in Eq.~\eqref{ScatteringTensor}.

Before addressing the integration, we construct a mapping of variables as $k_1 \rightarrow x_A ; k_2 \rightarrow x_B ; k_3 \rightarrow -x_C ; k_4 \rightarrow -x_D +G $, which makes the momentum Dirac delta completely symmetric with respect to each variable: $\delta(x_A+x_B+x_C+x_D)$. By doing so, the whole integral structure becomes symmetric with respect to all the variables. Thanks to this, we can, without loss of generality, invert the Dirac deltas with respect to the first two variables. The inversion with respect to any couple of variables can be obtained simply by an appropriate simple mapping (not shown here). 

In \ref{Appendix:Phi} we derive how to invert the Dirac deltas with respect to the first two variables to finally obtain an expression that has the following structure:
\begin{widetext}
\begin{equation}\label{MonteCarloIntegration}
\begin{split}
\Phi[ &F[],l_A,h_A,l_B,h_B,l_C,h_C,l_D,h_D,\mu_0,\mu_{A1},\mu_{A2},\mu_{B1},\mu_{B2},\mu_{C1},\mu_{C2},\mu_{D1},\mu_{D2}]=\int\displaylimits_{l_C} ^{h_C} \int\displaylimits_{l_D} ^{h_D} d x_C \;d x_D \\
&\Bigg( \frac{F[x_{A+}[x_C,x_D],x_{B+}[x_C,x_D],x_C,x_D]}{\sqrt{\mathcal{D}[x_C,x_D]}}. \Theta_{[l_A,h_A]}[x_{A+}[x_C,x_D]] \Theta_{[l_B,h_B]}[x_{B+}[x_C,x_D]]\Theta_{[0,\infty]}[\mathcal{D}[x_C,x_D]]+\\
& \frac{F[x_{A-}[x_C,x_D],x_{B-}[x_C,x_D],x_C,x_D]}{\sqrt{\mathcal{D}[x_C,x_D]}}. \Theta_{[l_A,h_A]}[x_{A-}[x_C,x_D]] \Theta_{[l_B,h_B]}[x_{B-}[x_C,x_D]]\Theta_{[0,\infty]}[\mathcal{D}[x_C,x_D]] \Bigg)
\end{split}
\end{equation}
\end{widetext}
where $\Theta_{[...,...]}[...]$ represents the Heaviside function between the edges (i.e.~$l_A$ and $h_A$ or $l_B$ and $h_B$) of the element corresponding to the variable reduced ($x_A$ or $x_B$ in this case), $\mathcal{D}$ is the discriminant of the quadratic equation obtained when reducing the energy Dirac delta, while the basis functions and $w$ in Eq.~\eqref{ScatteringTensor} are grouped in  F[...]. For the complete expressions of all the terms and a more accurate description of the integral, refer to \ref{Appendix:Phi}. Eq.~\eqref{MonteCarloIntegration} now does not have any internal Dirac deltas and requires an integration over a rectangular domain. In spite of the lack of Dirac deltas, the integrand has some strong discontinuities due to the presence of the Heaviside functions. In spite of the relatively low dimensionality of this integral in 1D, to be consistent with the treatment we will use in 2D and 3D materials, we use Monte Carlo to perform this integration.

The scattering tensor has some inherent symmetries, which are equivalent to particle, momentum and energy conservation. These symmetries would be automatically obeyed for an error-free calculation of the scattering tensor. However, numerical calculation introduces a finite error which in general breaks  these symmetries. In Ref.~\cite{Michael} we symmetrised the tensor after its calculation. We highlight here that, when an apposite construction of Monte Carlo points is done (not treated here), the symmetrisation of the scattering tensor elements becomes unnecessary, and those conservations are ensured from the outset regardless of the numerical precision.

\section{Time propagation algorithms}

Eq.~\eqref{CollisionIntegral} is a first order, semi-discrete (since the time variable has not been discretised yet) non-linear ordinary differential equation. In principle, once the scattering tensor for a particular scattering channel is calculated, the population can be propagated in time by simply contracting the scattering tensor with the populations at that time.  
To complete the discretisation, we need to use a time stepping algorithm (which is a numerical algorithm for solving initial value first order ordinary differential equations). We compare in this section two different time propagation schemes: Runge-Kutta 4 (RK4) and the adaptive time stepping method, Dormand Prince 853 (DP853).

\subsection{Runge-Kutta 4}

The classic RK4 is a fifth order accurate method. It is easy to implement and, more importantly, it requires only 4 function evaluations per step in time\cite{press2007numerical}.  RK4 is the most commonly used general purpose time stepping algorithm given that it combines ease of implementation, sufficient stability, good computational cost, a good order of convergence, and usually performs sufficiently well on a very large class of problems.

However, RK4 scheme also presents some serious limitations. The most critical for the present problem is that the timestep is fixed and needs to be provided by the user. This has two major drawbacks. Firstly, usually one would like to control precision (or tolerance to error) of the solution. With RK4 this must be done a posteriori by controlling the timestep. Moreover this must be done directly by the user, leading to an overuse of human time and decreased efficiency. For that reason, it would be beneficial to have a method that can decide the timestep itself.

However the most critical problem is that often dynamics can evolve through different timescales, with, for instance, a first timescale with relatively high time derivatives evolving towards a slower timescale with much smaller derivatives. This is the typical behaviour of the Boltzmann scattering operator, especially when several quasiparticles and bands are involved. This means that the error at a given step in time and, more importantly, the stability condition change throughout the dynamics. The fact that the timestep is fixed in RK4, means that it must be chosen according to the strictest restrictions, even if for the greatest part of the time propagation that chosen time step ends up being unnecessarily small. As a result the actual wall time of the simulations (the real computational cost) is inflated with no benefit on the precision.

RK4 still remains a powerful method in its simplicity, and we will use it to benchmark our results. Moreover in certain situations, especially when the system does not transition between different timescales, or (not shown here) in the presence of static electric fields, RK4 can still maintain a computational advantage over more advanced methods.

\subsection{Adaptive time step Dormand Prince 853}

Following the conclusions of the previous section we choose to implement an adaptive time stepping algorithm. These algorithms relieve the user from the task of setting the timestep, and, more importantly, allow for the timestep to be constantly adapted to convergence and stability requirements at each step. These algorithms usually work according to the following strategy. 1) They attempt a propagation with a given timestep $\Delta t$ (usually estimated from the previous steps). 2) They estimate the error. 3) They compare the error with the required accuracy, and if below it, they accept the step and advance the time by $\Delta t$, otherwise  the solution is discarded and restart from step 1 with a new estimation of $\Delta t$. 

Here we implement adaptive time stepping according to the Dormand Prince 853 method \cite{dormand1986runge} which is an eight order embedded Runge Kutta method. It uses 12 function evaluation per attempted step to both calculate the numerical solution and estimate the error. For details on the implementation see Ref.~\cite{DP853_Implementation}.

There is one last problem to address when using adaptive time stepping algorithms. Since the algorithm constantly adapts the timestep the output is constructed at non-uniformly spaced time values. This makes plotting and comparison with experiments or other simulations difficult. For this reason, all adaptive time step algorithms allow for interpolation of the solution between steps in time with high order accuracy. To distinguish this output from the direct output of the method, it is referred to as dense output. DP853 allows for the construction of a seventh order accurate dense output, yet achieving this accuracy between the adaptive timesteps requires further 3 function evaluations per step. Let us stress that dense output only serves to provide the user with interpolated values of the solution at the user-specified regular time intervals and that it does not affect the time stepping of DP853 in any way: DP853 uses its own previous time step solution to estimate the next time step value, not the dense output.

In spite of the vastly increased complexity of the solver compared to RK4, DP853 proves to be computationally advantageous compared to RK4 in the long run. 

\subsubsection{Modifications to the error estimation}

The version of DP853 that we have implemented follows very closely Ref.~\cite{DP853_Implementation}. However we have importantly modified the error evaluation to address some specific features of the Boltzmann scattering term. Before addressing the changes we have made, we first need to summarise the way typical Dormand Prince methods estimate the error. As there are quite some technicalities involved with DP853, and it is not our intention to repeat here a full description of this method, we will instead provide a brief description of the simpler Dormand Prince 5 (DP5) and show how we modify the error estimation in that case. The interested reader can easily apply our extension to DP853.

DP methods, being embedded RK methods, can estimate the solution at different orders of accuracy. DP5 begins the estimation of the error by taking the difference between two solutions at different orders
\begin{equation}
 	\left| \Delta_n(k) \right| =  \left| f_n^{[5]}(k) - f_n^{[4]}(k) \right| 
\end{equation} 
where $ \Delta_n(k)$ is the band- and momentum-resolved absolute value of the difference between the fifth-order RK estimation $f_n^{[5]}(k)$ and the fourth-order one $f_n^{[4]}(k)$. We want to compare the $k$-resolved difference with the so called scale which represents the local error that we are ready to tolerate. It is constructed by adding two terms: 
\begin{equation}
 	\mbox{scale}_n(k) = \mbox{atol}_n + \left| f_n^{[5]}(k) \right| \; \mbox{rtol}_n 
\end{equation}
The first term $\mbox{atol}_n $ is the absolute tolerance (which we allow to be band-dependent) and refers to the acceptable absolute error in the population. The second term gives the acceptable error as the product of a relative tolerance $\mbox{rtol}_n$ and the  value of the function (for which the highest order estimation is used). This allows for the error estimation to be controlled as a fraction of the actual population, unless such quantity becomes lower than the absolute tolerance, in which case the less strict requirement applies. The square of the normalised error is then estimated as
\begin{equation} \label{errorDP5}
 	\left( \epsilon_{\tiny\mbox{DP}5} \right)^2 =\frac{1}{N} \sum_n  \left( \int \left( \frac{ \left| \Delta_n(k) \right|}{ \mbox{scale}_n(k) } \right)^2 dk \Bigg/ \int dk \right),
\end{equation} 
where $N$ is the number of bands, and each integral  runs over each band's domain (even if the domain is not explicitly written for shortness). 

We can now introduce the first modification done to the error estimation. The relative tolerance term has the role of ensuring that the error is smaller than the information carried by the solution. Usually this information is the distance of the value of the solution from 0. However that is not necessarily always the case for fermionic populations. If the population is smaller than 0.5 then the relevant quantity is the number of electrons. However when the population is above 0.5, the system is better described by holes, and we want our precision to be compared to the number of holes i.e. ($1-f$). Therefore we want to make sure that the relative tolerance estimation works equally well for electrons and holes. We modify the scale  for fermionic bands as
\begin{equation}
 	\mbox{scale}_n(k) = \mbox{atol}_n + \min \left( \left| f_n^{[5]}(k) \right|,  \left| 1- f_n^{[5]}(k) \right| \right)\;  \mbox{rtol}_n 
\end{equation}
while no change is done for bosonic bands.

The second modification to the error estimation for DP853 is motivated by a different problem. The Boltzmann scattering term is an operator which has several fixed points. These fixed points include thermal equilibrium distributions (more fixed points exist, but it is not really relevant for this discussion). However that is true only when the populations acquire physically meaningful values: if the population somewhere is negative, or, for fermions, above 1, there is no guarantee that Eq.~\ref{ScatteringIntegral} will converge to a thermal equilibrium (yet it still might). If at any step in time the solution acquires unphysical values somewhere, the time evolution of Eq.~\ref{ScatteringIntegral} might lead to a complete divergence of the population. Notice that this would be still the legitimate time propagation of that initial condition and therefore DP has no way of recognising this behaviour by looking at the error. This means that an error, even within the acceptable tolerance, when leading to unphysical values of the solution, can cause catastrophic instabilities in the solution.

To prevent that behaviour we have to do an error estimation that does not treat all the errors of the same amplitude equally. An error that keeps the population within the window of physical values is treated in the standard way. On the other hand we artificially amplify any error that would lead the solution to acquire unphysical values. To achieve that, for each band, we add the following terms to the squared error in Eq.~\ref{errorDP5}:
\begin{equation}
 	\left(\epsilon_{n}\right)^2 = \mbox{p}_n\; \int \left( \frac{ \Theta\left(- f_n^{[5]}(k)\right)}{ \mbox{scale}_n(k) } \right)^2 dk \Bigg/ \int dk,
\end{equation} 
where $\mbox{p}_n$ is a penalisation factor (that we allow to be band dependent), and $\Theta$ is the unit step function. For fermionic bands we add the following further contribution to the squared error
\begin{equation}
 	\left(\epsilon_{{\tiny\mbox{ferm}},n} \right)^2= \mbox{p}_n\; \int \left( \frac{ \Theta\left(f_n^{[5]}(k)-1\right)}{ \mbox{scale}_n(k) } \right)^2 dk \Bigg/ \int dk
\end{equation} 
which penalises solutions with values above 1. DP, in its effort to contain the error by acting on the timestep, will then reject the step when these problematic cases arise, and reduce the timestep.

We remind that all the above expressions enter the code in their discretised versions (which we do not show here for brevity).

\section{Numerical results}

\begin{figure}
    \centering
    \includegraphics[width=0.45\textwidth]{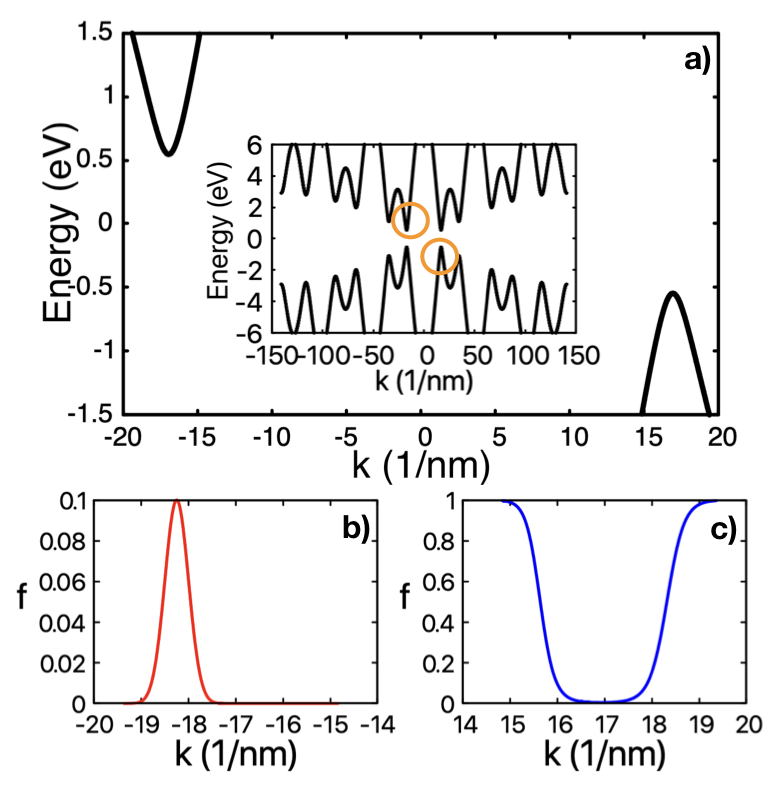}
    \caption {a) Band structure considered for the calculations done in this work. Inset: Band structure of CNT(6,5) calculated using tight binding \cite{malic2013graphene} from which the dispersions used in this work have been extracted. The circles mark the part of dispersion selected for this study. b) and c) Initial out-of-equilibrium distribution of electron population in the two bands. The chemical potential $\mu$ is set to -1 eV and the temperature value is set to 1000 for band 2 (c). Band 1 has been filled with a non equilibrium distribution (b) chosen to be a Gaussian centered at k=-18.25 with a spread of 0.25 and amplitude of 0.1 eV.} 
    \label{fig:InitialPopulation}
\end{figure}

We show numerical results obtained with the use of quadratic basis functions, and compare the performance of the two introduced time propagation schemes. It is not our intent here to address physically interesting cases, but only to show the capabilities and performance of the method on a minimal case study: a two band system. In view of applying the algorithm to the thermalisation dynamics of carbon nanotubes, we choose to describe a 1D semiconducting material. In order to highlight that the implementation can handle bands defined on different meshes, we describe an indirect bandgap semiconductor. Finally to show that the code works for arbitrary dispersions and how they can be taken from ab initio calculations, we use as dispersions those of two electronic bands of (6,5) carbon nanotubes (CNT) close to the Fermi level (see Fig.~\ref{fig:InitialPopulation}a) as calculated using tight-binding\cite{malic2013graphene}.  We include all possible electron-electron scattering channels obtainable with these two bands, as listed in table \ref{tab:ScatteringChannels}. Again it is not our purpose to describe a realistic system, so we choose all the scattering amplitudes, $w^{e-e}$, to be the same and equal to a constant 1, except for scattering channels 1, 3, and 5. For these the dependence on the momenta is chosen to be a constant 1 everywhere except in regions where the transferred momentum becomes smaller than 0.1. In that region the $w^{e-e}$ is chosen to be linear with the transferred momentum. This mimics the property of real scattering amplitude, which vanishes when the initial and final state are identical. Notice that enforcing this property is necessary as it is required to avoid a divergence of the scattering integral due to a divergence in the joint density of states. Nonetheless we stress that the choice of the dependence of the scattering matrix elements on the momenta is arbitrary and simply meant to display the capabilities of the method.

\begin{table}[tb]
    \centering
    \begin{tabular}{|c|c|}
        \hline
Scattering Channel number & Scattering Process \\ 
 \hline
 1 & 1+1 $\Longleftrightarrow$ 1+1 \\ 
 \hline
 2 & 1+1 $\Longleftrightarrow$ 1+2 \\ 
 \hline
 3 & 1+2 $\Longleftrightarrow$ 1+2 \\ 
 \hline
 4 & 2+2 $\Longleftrightarrow$ 2+1 \\ 
 \hline
 5 & 2+2 $\Longleftrightarrow$ 2+2 \\ 
 \hline
 6 & 1+1 $\Longleftrightarrow$ 2+2 \\ 
 \hline
    \end{tabular}
    \caption{List of all scattering channels included. The numbers on the left column refer to band numbers (1 for conduction band and 2 for valence band). For instance $2+2 \Longleftrightarrow 2+1$ describes a scattering channel where two electrons in band 2 scatter into band 1 and band 2.}
    \label{tab:ScatteringChannels}
\end{table}

\subsection{Time propagation: Runge Kutta 4}

\begin{figure}
    \centering
    \includegraphics[width=0.55\textwidth]{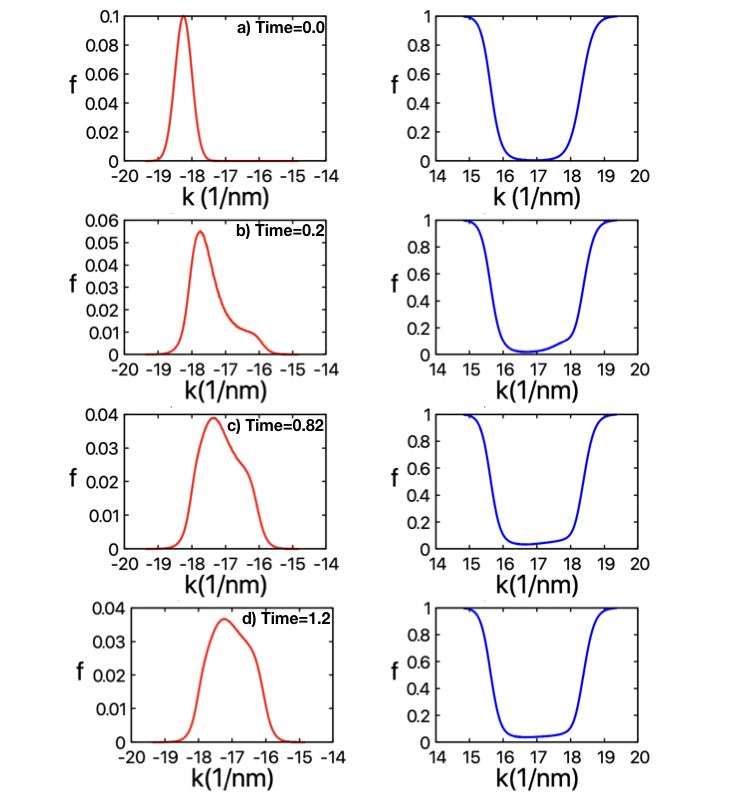}
    \caption {Time propagation of the initial electron population distribution in band 1 (left side) and band 2 (right side) with RK4 scheme. a), b), c) and d) show the different successive snap shots of the time propagation. Adaptive time stepping with DP853  shows the same thermalization profiles (not shown).} 
    \label{fig:Thermalization}
\end{figure}

To test the method we choose an out-of-equilibrium population as initial condition (see Fig.~\ref{fig:InitialPopulation} b and c) and let the code propagate the populations in all the bands. Notice that since the scattering amplitude is in arbitrary units, the time is in arbitrary units as well.
In Fig.~\ref{fig:Thermalization} we show the evolution of the populations. The calculations have been done with 100 elements per band (for a total of 600 basis functions), and a RK4 timestep of 0.001. We observe that the initial out-of-equilibrium distribution in the bands thermalises with time. To accommodate for the added particles and energy (the initial gaussian excitation in the higher energy band) the thermalized distribution broadens in both bands as seen from Fig.~\ref{fig:Thermalization}. The shift of the peak in band 1 towards the center of the domain indicates a re-distribution of the momentum between the two bands. Notice that when umklapp scatterings are present, total momentum is not preserved, as these scatterings  preserve it only up to a reciprocal lattice vector $G$.  Since we constructed the Brillouin zone to be much wider than the momentum domain of the bands, umklapp scatterings (which the solver tries to construct by default) are absent. This implies that the total momentum is conserved to machine precision. We have also verified that the code conserves total particles and total energy to machine precision.

\begin{figure}
    \centering
    \includegraphics[width=0.45\textwidth,height=6cm]{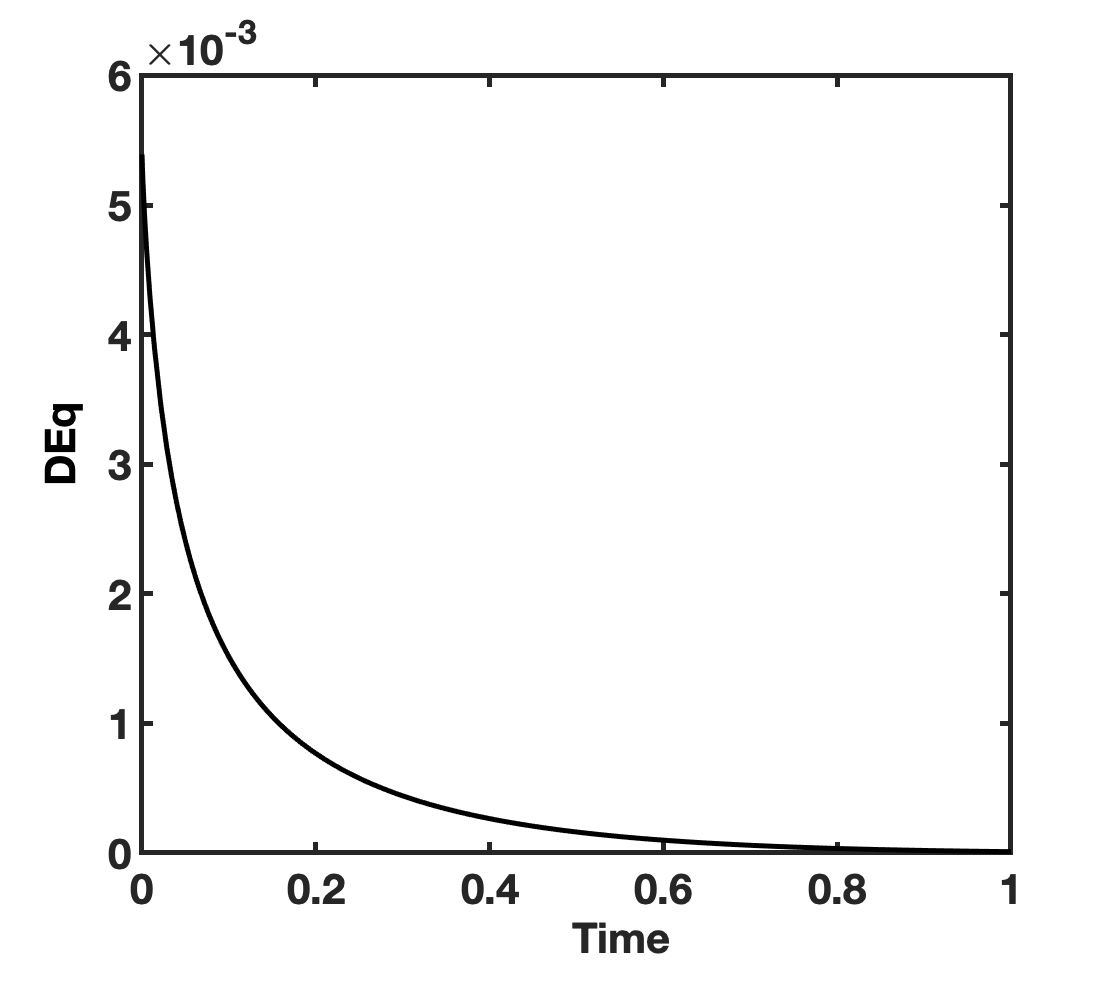}
    \caption {Distance from equilibrium (as defined in Eq.~\ref{eq:distanceEquilibrium}) for the time propagation of the excitation shown in Fig.~\ref{fig:InitialPopulation} using RK4 scheme. The band index is omitted for brevity.} 
    \label{fig:DistanceFromEquilibrium}
\end{figure}

Before proceeding to the numerical evaluation of the method, as an interesting analysis tool we define a quantity that indicates the distance from equilibrium. We define the distance between two populations $f(k)$ and $g(k)$ in a given band as
\begin{equation} \label{eq:distanceEquil}
 	d(f(k),g(k)) =\left( \int \left(f(k) - g(k) \right)^2 dk \right)^{1/2}
\end{equation}
and denote the band-resolved distance from equilibrium $DEq_n$ as the distance between the population $f_n(k,t)$ of that band $n$ at any given time $t$
\begin{equation} \label{eq:distanceEquilibrium}
    DEq_n =d(f_n(k,t),f_{eq,n}(k))  
\end{equation}
and the equilibrium condition $f_{eq,n}(k) = \lim_{t \to \infty} f_n(k,t)$. The distance from equilibrium for the conduction band (as shown in Fig.~\ref{fig:DistanceFromEquilibrium}) is not a simple exponential decay. 

\begin{figure}
    \centering
    \includegraphics[width=0.45\textwidth]{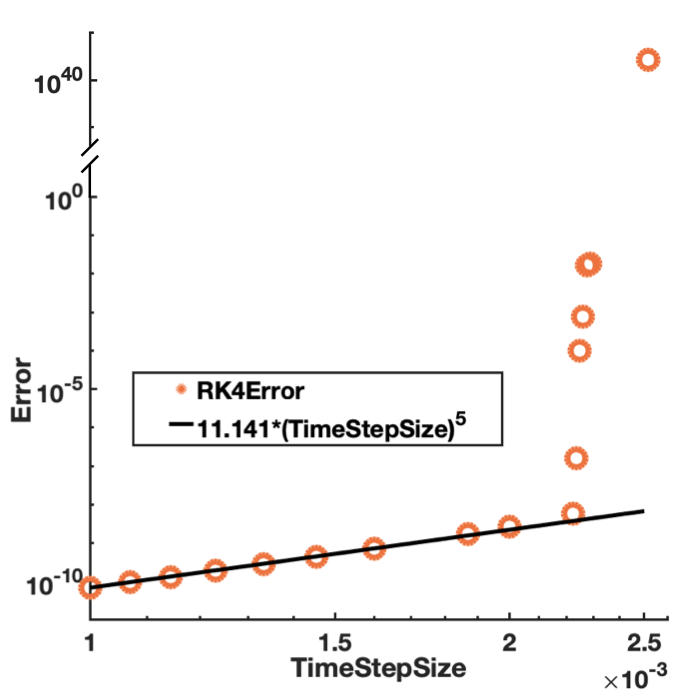}
    \caption {Variation of Error with the timestep size for RK4. For smaller time steps the error varies with the fifth power of the time step size.} 
    \label{fig:RK4Error}
\end{figure}

We test here the dependence of the precision of RK4 on the timestep. To construct the error, we need the exact solution. As that is not accessible, we perform a DP853 calculation with a very small absolute and relative tolerance (1e-16 for both the bands), and take this solution as a good estimation of the exact solution. We then perform a series of RK4 calculations with increasingly larger timesteps and calculate the distance between the predicted population at time $t=0.4$ with our nearly exact solution at $t=0.4$. In Fig.~\ref{fig:RK4Error} we show the influence of the time step on the error.

\begin{figure}
    \centering
    \includegraphics[width=0.45\textwidth]{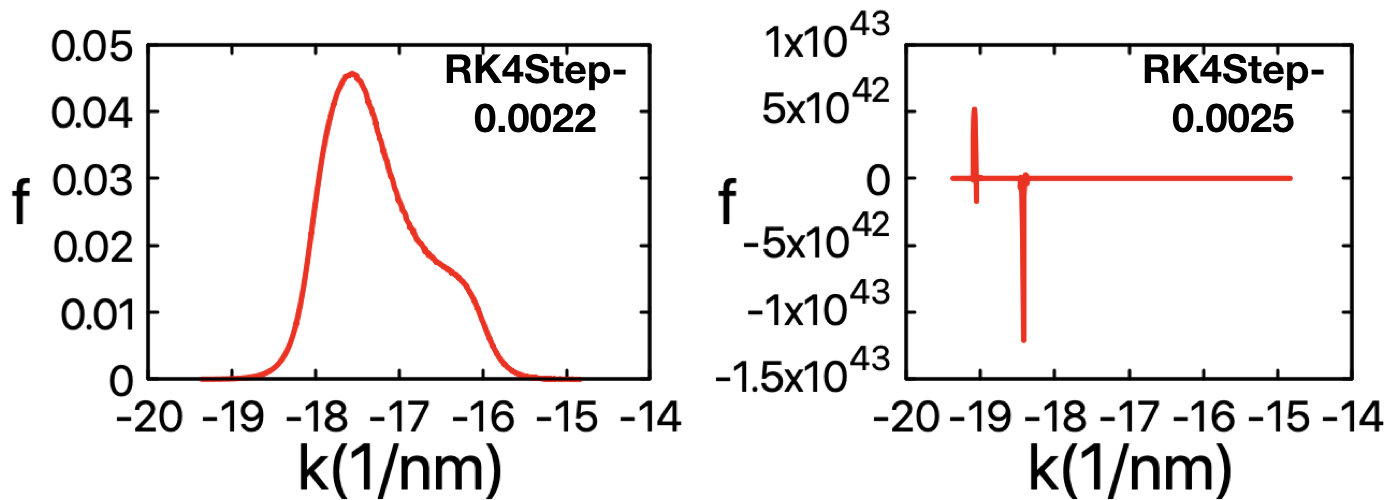}
    \caption {Effect of instability on the population. Right) Stable solution at time $t=0.4$. Left) Solution at time $t=0.4$ when the RK4 time propagation becomes unstable.} 
    \label{fig:popInstab}
\end{figure}

We observe that with increasing timestep the error increases as $\delta t^5$ as expected. We can also observe the appearance of instability, when the timestep increases above a certain threshold. In Fig.\ref{fig:popInstab} we show the population for a timestep of $0.0022$ where instability in the solution is just appearing, and the one for a timestep of $0.0025$ where instead the solution has completely diverged.

\subsection{Time propagation: Dormand Prince 853}

The observation that the system's thermalisation evolves through different timescales (Fig.~\ref{fig:DistanceFromEquilibrium}), hints that a full propagation with a single timestep is inefficient. Early times require small timesteps, while later times could permit the use of longer ones. Even if that can be done by the user manually in RK4 by stopping the simulation, changing the timestep and then continuing, adaptive time stepping methods like DP853 are preferable in these cases, as they perform that task automatically at each timestep and optimise the choices.

\begin{figure}
    \centering
    \includegraphics[width=0.45\textwidth]{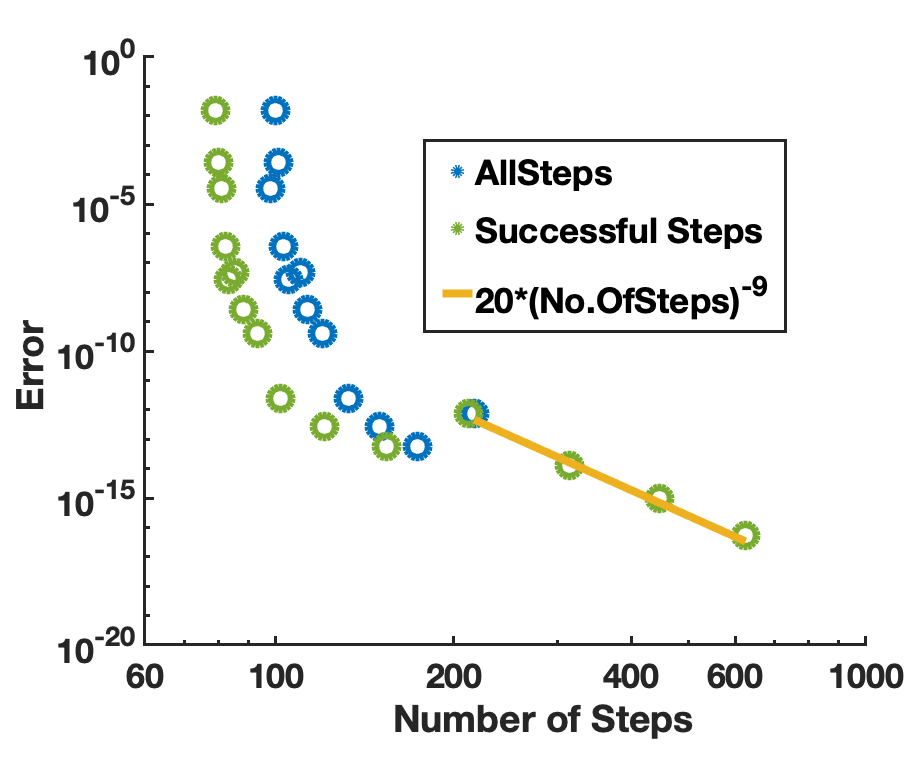}
    \caption {Variation of Error for DP853 with the number of steps. At higher number of steps, i.e.~higher tolerances, the error varies with the ninth power of the number of the time steps.} 
    \label{fig:DP853Nsteps}
\end{figure}

The analysis of the convergence order of DP853 itself is less straightforward compared to RK4, as there is no fixed timestep. Moreover the computational cost is not directly linked to the number of performed steps, since a fraction of the steps are rejected. We first show in fig.~\ref{fig:DP853Nsteps} the behaviour of the error in the solution after a time, t=0.4 in dependence of the accepted and total (meaning accepted plus rejected) steps in time. Both values are indirectly controlled by setting the tolerances mentioned in the previous section. We observe that the error scales with the ninth power of the total timesteps at lower tolerances or higher number of time steps. It can be noted from Fig.~\ref{fig:DP853Nsteps} that the number of rejected timesteps (i.e.~the difference between the number of total timesteps and the number of successful timesteps) tends to remain relatively constant over the full range of required tolerances (please notice the logarithmic scale). We observe that most of the rejections happen at the beginning of the calculation. DP853 uses the error obtained using the initial guess for timestep (which we provide as large) to make a better guess. This takes a few iterations before DP853 obtains a value of the time step that gives an acceptable error. From that point on DP853 is very efficient in predicting a time step for the next iteration and rarely rejects future steps.

\begin{figure}
    \centering
    \includegraphics[width=0.45\textwidth]{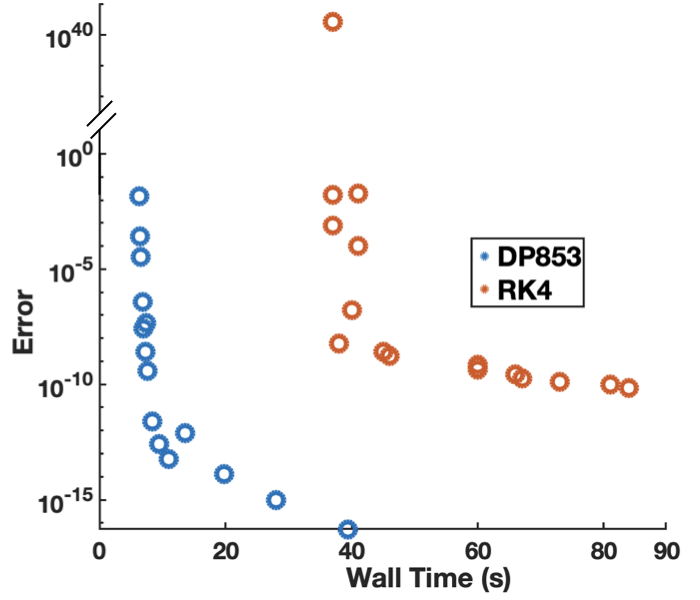}
    \caption {Comparison of the computational cost of DP853 and RK4.} 
    \label{fig:ErrorComp}
\end{figure}

We can now compare DP853 and RK4. The number of total steps taken does not provide a fair comparison, since the computational cost at each step is very different for the two methods. We therefore compare in Fig.~\ref{fig:ErrorComp} the wall time (actual time that it takes for the time propagation) at parity of error. As seen from Fig.~\ref{fig:ErrorComp} DP853 vastly outperforms RK4. This is partially due to the much higher convergence order. However notice that RK4 diverges if the numerical effort is too low. This is due to the fact that while DP853 adapts its step and can take shorter time steps at early times when the dynamics is fast and then save computational resources later, RK4 does not have this flexibility and its overall stability is linked to the worst case timescale.  For a required error precision of 1e-4, which can be considered acceptable, the cost of RK4 is approximately 45 times the cost of DP853. 

However Fig.~\ref{fig:ErrorComp} does not account for one of the greatest advantages of DP853 over RK4. In RK4 the user needs to test timestep sizes and observe the solution to evaluate its quality. This approach has both a numerical and, even more importantly, a human cost, not included in the comparison in Fig.~\ref{fig:ErrorComp}. This makes the real wall time for RK4 much higher than the one for DP853, in almost all the cases.

\subsection{Dormand Prince 853: Dense output}

\begin{figure}[h!]
    \centering
    \includegraphics[width=0.5\textwidth]{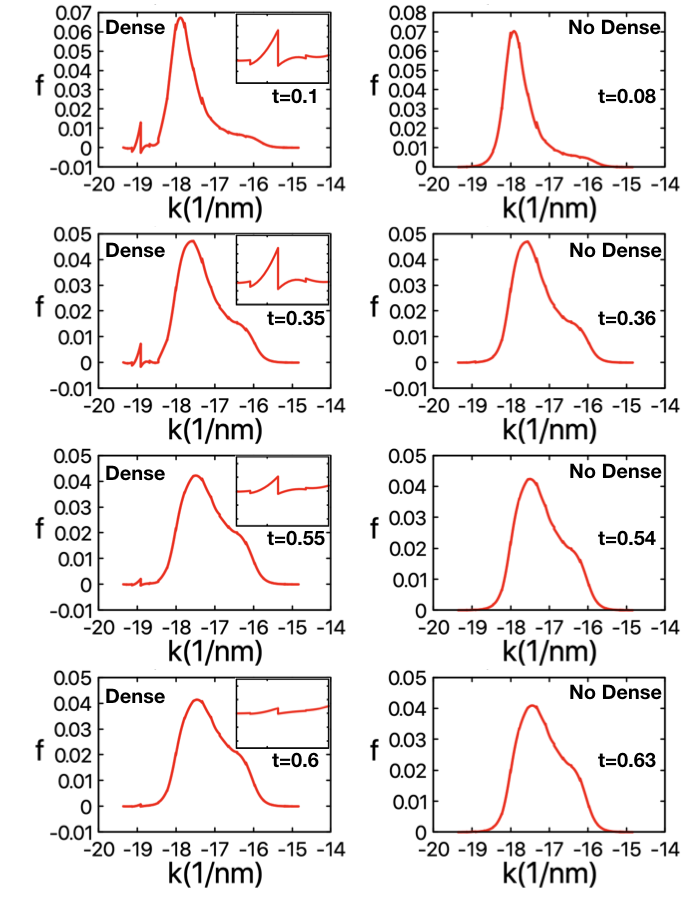}
    \caption {Comparison between the eight-order solution (on the adaptive time steps) and the seventh-order solution (on the closest point of the dense mesh) for DP853. Inset: Zoomed in discontinuity of the solution. To magnify the effect of numerical accuracy we choose a lower mesh resolution of n=20 elements and set the tolerance values to 1e-2 for this case.} 
    \label{fig:Dense}
\end{figure}

It is worth however highlighting one problem we encountered when using DP853. The method computes the solution with eight-order accuracy at irregular time intervals. However this output, due to this characteristic, is seldom useful. The order of convergence of the dense output is one order lower. That means that while DP853 controls the error of the eight-order embedded Runge Kutta method by comparing it to the tolerances assigned by the user, the useful output (the dense one) has a (sometimes importantly) higher error.  This difference in the orders of convergence can lead to very evident errors. In Fig.~\ref{fig:Dense} we show one case where the eight-order solution at the adaptive steps has an extremely low error, while the seventh-order one at the dense time mesh shows evident errors. One might see this from a different point of view: while the solution produced on the adapted times has a high precision, the lower order interpolation, used to produce the solution on the dense mesh has a lower precision. This is an issue the user should be aware of, yet it is easily solvable by running another simulation with smaller tolerances with the effect of reducing the error in both orders. This is rarely a problem, since, given the high order of convergence, usually an even rather large decrease in the required tolerance leads to a very small increase in computational time.

\section{Conclusions}

Concluding, we have extended in two major ways the method we developed in Ref.~\cite{Michael} for the solution (without any approximation) of the time-dependent Boltzmann scattering integral for strongly-out-of-equilibrium scenarios and for realistic band structures and matrix elements. 1) We have extended the treatment to higher order basis functions, gaining one order of convergence in momentum space. 2) We have implemented a powerful adaptive time step scheme and shown how to modify it to address the issues typical of this equation. We have shown how these improvements work and allow for flexible and multipurpose calculations of the ultrafast time propagation of femtosecond excitations in solids.

\appendix
\gdef\thesection{Appendix \Alph{section}}
\renewcommand\theequation{\Alph{section}.\arabic{equation}}

\section{Scattering elements with second order basis functions}\label{Appendix:Phi}

We here address the task of computing the scattering elements in Eq.~\ref{ScatteringTensor}. As already mentioned we first perform the mapping of variables as: $k_1 \rightarrow x_A ; k_2 \rightarrow x_B ; k_3 \rightarrow -x_C ; k_4 \rightarrow -x_D +G$, to express the integral in a form that is symmetric with respect to the variables. It can be shown that the integral assumes the following general form:
\begin{widetext}
\begin{equation}\label{AppendixEq:PhiMapped}
    \begin{split}
     \Phi&[F(),l_A,h_A,l_B,h_B,l_C,h_C,l_D,h_D,\mu_0,\mu_{A1},\mu_{A2},\mu_{B1},\mu_{B2},\mu_{C1},\mu_{C2},\mu_{D1},\mu_{D2}] \\
&= \int_{l_A}^{h_A}\int_{l_B}^{h_B}\int_{l_C}^{h_C}\int_{l_D}^{h_D}  dx_A dx_B dx_C dx_D\; F\left(x_A,x_B,x_C,x_D\right)\;\delta(x_A+x_B+x_C+x_D)\\
& \delta(\mu_0+\mu_{A1}x_A+\mu_{A2}x_A^2+\mu_{B1}x_B+\mu_{B2}x_B^2+\mu_{C1}x_C+\mu_{C2}x_C^2+\mu_{D1}x_D+\mu_{D2}x_D^2)
    \end{split}
\end{equation}
\end{widetext}
where the function $F()$ incorporates the basis functions and the scattering matrix element, while the coefficients, $\mu_0$, $\mu_{A1}$ etc.~result from expressing the coefficients of the dispersions and the basis functions, everything after the appropriate variable transformation.

Without loss of generality, we now choose the variables $x_A$ and $x_B$ to analytically invert the Dirac deltas. By reducing the first Dirac delta, we can write the first variable as 
\begin{equation}
 	x_A[x_B,x_C,x_D]=-x_B-x_C-x_D.
\end{equation}
We can now substitute this in the energy Dirac delta and collect all the terms with the same power of $x_B$ to find:
\begin{equation}\label{AppendixEq:QuadDelRed}
    \delta\left(energy\right)=\delta\left(\alpha x_B^2+\beta [x_C,x_D] x_B+\gamma [x_C,x_D]\right)
\end{equation}
where
\begin{align}
 &\alpha=\mu_{A2}+\mu_{B2}\\
&\beta [x_C,x_D]=\mu_{B1}-\mu_{A1}+2\mu_{A2}(x_C+x_D)\\
& \gamma [x_C,x_D]= \mu_0+(\mu_{C1}-\mu_{A1})x_C+ \\
&\;\;\;\;\;\;\;\;\; + (\mu_{A2}+\mu_{C2})x_C^2+ +(\mu_{D1}-\mu_{A1})x_D+ \nonumber \\
&\;\;\;\;\;\;\;\;\;+(\mu_{A2}+\mu_{D2})x_D^2+2\mu_{A2}x_Cx_D \nonumber
\end{align}

The quadratic form inside the Dirac delta in Eq.~\ref{AppendixEq:QuadDelRed}, possesses two roots with respect to the variable $x_B$. Notice that these roots $x_{B+}[x_C,x_D]$ and $x_{B-}[x_C,x_D]$ are dependent on the last two variables and are real only where the discriminant 
\begin{equation}
 	D[x_C,x_D]=\beta^2[x_C,x_D] - 4\alpha[x_C,x_D] \gamma[x_C,x_D]
\end{equation}
is positive. We can now rewrite Eq.~\ref{AppendixEq:PhiMapped} as 
\begin{widetext}
\begin{equation}\label{AppendixEq:FinalIntform}
    \begin{split}
\Phi[...]=\int_{l_C}^{h_C}\int_{l_D}^{h_D}d x_C d x_D \Bigg( &\frac{F[x_{A+}[x_C,x_D],x_{B+}[x_C,x_D],x_C,x_D]}{\sqrt {D[x_C,x_D]}} \cdot\\
&\cdot \Theta_{[l_A,h_A]}[x_{A+}[x_C,x_D]]\;\; \Theta_{[l_B,h_B]}[x_{B+}[x_C,x_D]] \;\; \Theta_{[0,\infty]}[D[x_C,x_D]]\\
&+ \frac{F[x_{A-}[x_C,x_D],x_{B-}[x_C,x_D],x_C,x_D]}{\sqrt {D[x_C,x_D]}} \cdot\\
& \cdot\Theta_{[l_A,h_A]}[x_{A-}[x_C,x_D]] \;\; \Theta_{[l_B,h_B]}[x_{B-}[x_C,x_D]] \;\; \Theta_{[0,\infty]}[D[x_C,x_D]]\Bigg),
\end{split}
\end{equation}
\end{widetext}
where the first (second) addend in the brackets comes to the first (second) root. The $\Theta[...]$ represents the Heaviside function. The first two Heaviside functions derive from the limits of the integration in the first two variables. The third $\Theta[...]$ function, ensures that the integral is not performed over complex solutions. 

The expression in Eq.~\ref{AppendixEq:FinalIntform} is now a lower dimensional integral finally over a function and not anymore a distribution (in the sense of a generalised function). However this is not a smooth function due to the presence of the Heaviside functions. We are now free to perform this integral by any numerical technique. For consistence to the technique that we will use in higher dimensions, we performed this integral with Monte Carlo.

\bibliographystyle{unsrt}
\bibliography{References}

\end{document}